\documentclass[
 aip,
 jcp,
 amsmath,amssymb,
 reprint,
]{revtex4-1}

\usepackage{graphicx}
\usepackage{dcolumn}
\usepackage{bm}

\usepackage[utf8]{inputenc}
\usepackage[T1]{fontenc}
\usepackage{mathptmx}
\usepackage{etoolbox}

\makeatletter
\def\@email#1#2{%
 \endgroup
 \patchcmd{\titleblock@produce}
  {\frontmatter@RRAPformat}
  {\frontmatter@RRAPformat{\produce@RRAP{*#1\href{mailto:#2}{#2}}}\frontmatter@RRAPformat}
  {}{}
}%
\makeatother

\newcommand{\mybar}[1]{\mkern 1.5mu\overline{\mkern-1.5mu#1\mkern-1.5mu}\mkern 1.5mu}
\usepackage{upgreek}
\newcommand{\vc}[1]{\bm{#1}}
\DeclareMathAlphabet{\mathsfit}{T1}{\sfdefault}{\mddefault}{\sldefault}
\SetMathAlphabet{\mathsfit}{bold}{T1}{\sfdefault}{\bfdefault}{\sldefault}

\newcommand{\ud}{\ensuremath{\mathrm{d}}}
\newcommand{\uDelta}{\ensuremath{\Delta}}
\newcommand{\ui}{\ensuremath{\mathrm{i}}}
\newcommand{\ue}{\ensuremath{\mathrm{e}}}
\newcommand{\upi}{\ensuremath{\pi}}

\newcommand{\sumM}{\sum_{n=-\mybar{M}}^{\mybar{M}}}

\newcommand{\aren}{a_{\mathrm{ren}}}
\newcommand{\omegac}{\omega_{\mathrm{c}}}
\newcommand{\omegaalpha}{\omega_\alpha}
\newcommand{\omegan}{\omega_n}
\newcommand{\omegaalphan}{\omega_{\alpha n}}
\newcommand{\nbath}{n_{\mathrm{b}}}
\newcommand{\malpha}{m_\alpha}
\newcommand{\xalpha}{x_\alpha}
\newcommand{\galpha}{g_\alpha}
\newcommand{\eqn}[1]{Eq.~(\ref{#1})}
\newcommand{\Eqn}[1]{Equation~(\ref{#1})}

\newcommand{\eqnn}[2]{Eqs.~(\ref{#1}) and (\ref{#2})}

\newcommand{\eqnt}[2]{Eqs.~(\ref{#1})--(\ref{#2})}

\newcommand{\diff}[2]{\frac{\ud{}#1}{\ud{} #2}}

\usepackage{xcolor}

\begin{document}

\preprint{AIP/123-QED}

\title{Comparison of Matsubara dynamics with exact quantum dynamics for an oscillator coupled to a dissipative bath}
\author{Adam Prada}
\author{Eszter S.\ Pós}%
\altaffiliation[Current address: ]{MPI for the Structure and Dynamics of Matter, Luruper Chaussee 149, 22761 Hamburg, Germany}
\author{Stuart C.\ Althorpe}
\email{sca10@cam.ac.uk}
\affiliation{Yusuf Hamied Department of Chemistry, University of Cambridge, Lensfield Road, Cambridge, CB2 1EW, UK}

\date{\today}

\begin{abstract}
Matsubara dynamics is the classical dynamics which results when imaginary-time path-integrals are smoothed; it conserves the quantum Boltzmann distribution and appears in drastically approximated form in path-integral dynamics methods such as (thermostatted) ring-polymer molecular dynamics (T)RPMD and centroid molecular dynamics (CMD). However, it has never been compared directly with exact quantum dynamics for non-linear operators, because the difficulty of treating the phase has limited the number of Matsubara modes to fewer than 10. Here, we treat up to $\sim$200 Matsubara modes in simulations of a Morse oscillator coupled to a dissipative bath of harmonic oscillators. This is done by expressing the Matsubara equations of motion in the form of a generalised Langevin equation, approximating the noise to be real, and analytically continuing the momenta to convert the Matsubara phase into ring-polymer springs. The resulting equations of motion are stable up to a maximum value of modes which increases with bath coupling strength and decreases with system anharmonicity.  The dynamics of the tail of highly oscillatory Matsubara modes is found to be harmonic, and can thus be computed efficiently.  For a moderately anharmonic oscillator with a strong but subcritical coupling to the bath, the Matsubara simulations yield non-linear $\large\langle{\hat q^2\hat q^2(t)}\large\rangle$ time-correlation functions in almost perfect agreement with the exact quantum results. Reasonable agreement is also obtained for weaker coupling strengths, where errors arise because of the real-noise approximation. These results give strong evidence that Matsubara dynamics correctly explains how classical dynamics arises in quantum systems which are in thermal equilibrium.
\end{abstract}

\maketitle
\section{Introduction}

Matsubara dynamics was introduced in ref.~\onlinecite{Hele2015} as a hypothesis to explain how classical dynamics can coexist with quantum Boltzmann statistics. The results of `classical dynamics--quantum statistics' simulation methods \cite{Miller2001,Liu2011,Liu2015,Liu2009,Shi2003a,Poulsen2003,Craig2004,Craig2005,Craig2005a,Miller2005,Habershon2013,Rossi2014,Habershon2013,Markland2018,Cao1994,Cao1994c,Medders2015,Reddy2017,Trenins2019,Benson2019,Althorpe2021} suggest that Born-Oppenheimer nuclear dynamics in the condensed phase often lives in such a regime. For example, zero-point energy shifts in the infrared spectrum of liquid water are mainly corrected by including quantum (Boltzmann) effects into the statistics,\cite{Habershon2008, Habershon2009, Habershon2009a,Medders2015,Reddy2017} and rate coefficients are dominated by the quantum Boltzmann distribution at the barrier top.\cite{Craig2005,Craig2005a,Miller2001,Liu2009,Richardson2009,Richardson2016,Richardson2018a}

Standard theory, in the guise of the classical Wigner or LSC-IVR method,\cite{Miller2001,Liu2011,Liu2015,Liu2009,Shi2003a,Poulsen2003}
can only account for the classical dynamics--quantum statistics regime in the short-time limit, since it follows the classical dynamics of localised particles, which do not conserve the quantum Boltzmann distribution.
Matsubara dynamics, by contrast, follows the classical dynamics of smooth imaginary-time path integrals (in their own extended phase-space). Surprisingly, it is the smoothness alone that makes the dynamics classical,  and also makes it conserve the quantum Boltzmann distribution.\cite{Hele2015}

Matsubara dynamics cannot be used as a method (because of a serious phase problem), but it has proved useful at understanding the strengths and weaknesses of the heuristic path-integral methods [thermostatted] ring-polymer molecular dynamics ([T]RPMD)\cite{Craig2004,Craig2005,Craig2005a,Miller2005,Habershon2013,Rossi2014} and centroid molecular dynamics (CMD),\cite{Cao1994,Cao1994c,Medders2015,Reddy2017} which were shown to be (respectively) short-time and mean-field approximations to Matsubara dynamics.\cite{Hele2015a} This finding led to the quasi-centroid molecular dynamics (QCMD) method of Trenins et al.\cite{Trenins2019,Haggard2021} for simulating infrared spectra (which in turn has led more recently to the fast QCMD (f-QCMD)\cite{Fletcher2021} and T-PIGS methods\cite{Musil2022}). Perturbative applications of Matsubara dynamics have also explained why classical molecular dynamics (MD), CMD and RPMD underestimate the intensities of overtone and combination bands by typically an order of magnitude.\cite{Benson2021,Ple2021}

Given these and other potential uses of Matsubara dynamics, it would be good if one could at least simulate it numerically for some model benchmark systems, in order to confirm that it does correctly describe the classical dynamics--quantum statistics regime. This was done by Trenins et al.\ for the `Champagne Bottle' model,\cite{Trenins2018} where spurious red shifts\cite{Ivanov2010, Witt2009} in the CMD spectrum were shown to arise from neglect of the Matsubara modes describing dynamical fluctuations of the imaginary time-path Feynman paths around its centroid; when these modes were included, the red shifts vanished, bringing the spectrum into good agreement with the exact quantum result. However, in these simulations, the (dipole moment) operators were linear, and an artificial window function was used to damp the time.  

Non-linear operators are much more difficult to treat because they depend explicitly on the dynamics of the non-centroid modes. From static Matsubara calculations, it is known that the number of explicit modes converges very slowly, and that this is caused by the smoothness of the paths: non-smooth paths (i.e.\ the usual `ring-polymers' used in standard path-integral methods) converge much more rapidly,\cite{Ceperley1995,Coalson1986,Doll1999,Chakravarty1998} such that, say, 100 ring-polymer `beads' would give the same numerical convergence as 10\,000 Matsubara modes. The highly oscillatory modes have small amplitudes and thus can be expected to behave harmonically, allowing the use of a `tail' correction (as has been done in static calculations\cite{Coalson1986,Doll1999}). However, this would still leave many Matsubara modes needing to be treated numerically, each of them contributing to the extremely oscillatory Matsubara phase. 

One possible way to remove the phase was suggested in ref.~\onlinecite{Hele2015a}. This is to analytically continue the Matsubara momenta, which converts the phase into the smoothed equivalent of the familiar `ring-polymer' distribution that is sampled in standard path-integral methods. Unfortunately, this transformation pushes the dynamics into the complex plane, meaning that trajectories become unstable after very short times. Thus no analytically continued Matsubara calculations have been attempted to date. 

In this article, we show that analytically continued Matsubara dynamics is stable when the system is coupled sufficiently strongly to a bath of harmonic oscillators. The bath modes can be analytically continued without introducing instability, giving rise to a generalised Langevin equation (GLE), in which each Matsubara mode of the system is coupled to its own bath of Matsubara oscillators. The GLE has complex noise, but the noise can be made real using a decorrelation approximation introduced by Ivanov et al.,\cite{Karsten2018} which has only a minor effect on the calculated time-correlation functions for a sufficiently strong bath strength. The real-noise GLE is then found to stabilise the analytically continued (system) Matsubara trajectories, such that up to 200 Matsubara modes can be included in the dynamics. 
After summarising Matsubara dynamics and the system-bath model in Sec.~II, we derive the Matsubara GLE with complex and real noise in Sec.~III, then test its numerical stability and compare with numerically exact hierarchical equations of motion (HEOM)\cite{Tanimura1989,Tanimura1991,Tanimura2020,Shi2011,Zhu2012} results in Sec.~IV; Sec.~V concludes the article.

\section{Background theory}
\subsection{Matsubara dynamics}
To derive Matsubara dynamics,\cite{Hele2015,Trenins2018,Althorpe2021} one starts with the exact quantum time-correlation function (TCF). The derivation follows most naturally from the Kubo-transformed TCF,\cite{Kubo1957} 
\begin{equation}
    \widetilde{C}_{AB}(t) = \frac{1}{Z} \frac{1}{\beta}\int_{0}^{\beta}\ud \lambda\ \mathrm{Tr}\left[\ue^{-(\beta-\lambda)\hat{H}}\hat{A}\ue^{-\lambda\hat{H}}\ue^{i\hat Ht/\hbar}\hat{B}\ue^{-i\hat Ht/\hbar}\right].
    \label{qexact}
\end{equation}
where $Z$ is here the quantum partition function (and we will use $Z$ in what follows as a general symbol to denote the partition function and any multiplicative constants that are required to normalise whatever distribution appears in the integral). One expresses the quantum
dynamics as an (exact) propagation of the imaginary-time Feynman paths, then makes the approximation that these paths are smooth loops in imaginary time $\tau=0\rightarrow\beta\hbar$. Keeping in one-dimension to simplify the notation (generalisation to multi-dimensions is straightforward), these loops can be written as periodic functions ${p}_t(\tau),{q}_t(\tau)$ of imaginary time. In practice, one expands ${q}_t(\tau)$ (and similarly ${p}_t(\tau)$) as a Fourier series in terms of $M$ Matsubara modes ${Q}_n(t)$, as 
\begin{align}
{q}_t(\tau) = {Q}_0(t) + \sqrt{2}\sum_{n=1}^{\overline M}{ Q}_n(t)\sin(\omega_n \tau) + {Q}_{\mybar{n}}(t)\cos(\omega_n \tau) 
\end{align}
where $\overline{M}=(M-1)/2$ (so $M$ is odd), \(\mybar{n} \equiv -n\), and $\omegan = 2 \upi n/\beta \hbar$ are the Matsubara frequencies.\footnote{Note that $\omegan$ is negative for negative $n$.}

 The effect of this smoothing is to strip out all real-time quantum coherences, so that the dynamics of $({p}_t(\tau),{q}_t(\tau))$ becomes classical, with equations of motion $\dot { q}_t(\tau)={ p}_t(\tau)/m$, $\dot { p}_t(\tau)=-\ud V[{ q}_t(\tau)]/\ud{ q}_t(\tau)$. In practice, one evaluates this dynamics in the space of the Matsubara modes  ${P}_n(t),{Q}_n(t)$, using the Hamiltonian   \begin{equation}
H_M\left(\vc{Q},\vc{P}\right) = \sumM \frac{P_n^2}{2m} + U_M\left(\vc{Q}\right).
\end{equation}
where the potential of mean force  $U_M\left(\vc{Q}\right)$ is given in Appendix~\ref{app:mean_force}.\footnote{Two derivations of Matsubara dynamics have been reported, which smooth the imaginary-time Feynman paths in different ways: ref.~\onlinecite{Hele2015} smooths the paths by truncating the ring-polymer normal modes; ref.~\onlinecite{Trenins2018} by  Boltzmann-averaging over the modes $|n|>\overline{M}$. We use 
the latter approach here.}
The Matsubara dynamics approximation to the quantum Kubo TCF\ is the $M\to\infty$ limit of 
\begin{multline}
\widetilde{C}_{AB}^{\ [M]}(t) = \frac{1}{Z}\int\ud \vc{Q}\,\ud \vc{P}\,
 \ue^{-\beta H_M\left(\vc{Q},\vc{P}\right)}\ue^{\ui\beta \theta_M\left(\vc{Q},\vc{P}\right)}\\ \times A\left(\vc{P},\vc{Q}\right)B\left[\vc{P}(t),\vc{Q}(t)\right],
\label{eq:mats_tcf}
\end{multline}
where
\begin{equation}
\theta_M\left(\vc{Q},\vc{P}\right) = \sumM \omega_n P_n Q_{\mybar{n}},
\end{equation}
is the Matsubara phase.

The smoothness of $({ p}_t(\tau),{q}_t(\tau))$ ensures that the dynamics of $({ P}_n(t),{ Q}_n(t))$ conserves the phase
$\theta_M$ (because  $\theta_M$ is the angular momentum conjugate to the internal rotation of the loop, on which there is no torque), and thus conserves the quantum Boltzmann distribution in \eqn{eq:mats_tcf}. Unfortunately
$\exp(-\beta\theta_M)$ is highly oscillatory, which makes it numerically extremely difficult to sample the distribution in \eqn{eq:mats_tcf}. This is the `phase problem' mentioned in the Introduction, which we remove for a system-bath model by analytic continuation below (Sec.~III).

When $\hat A$ and $\hat B$ in \eqn{qexact} are linear functions of $\hat{{p}}$ and $\hat{{q}}$,  the functions
$A\left(\vc{P},\vc{Q}\right)$ and $B\left[\vc{P}(t),\vc{Q}(t)\right]$ in \eqn{eq:mats_tcf} become linear functions of the centroids, ${P}_0$ and ${Q}_0$ (i.e.\ the centres of mass of the loops $({p}_t(\tau),{q}_t(\tau))$). This makes Matsubara dynamics much less numerically intractable (though still very difficult) for linear operators, since one only needs to know explicitly the Matsubara dynamics of the centroids, which often allows one to ignore the dynamics of all but a few of the $n\ne 0$ modes describing dynamical fluctuations around the centroids.
  This property was exploited by Trenins et al. when computing the spectrum of the Champagne Bottle model,\cite{Trenins2018} and is also the basis of the practical path-integral methods RPMD, CMD and QCMD. The aim of this article, however, is to deal with non-linear operators, which require that one calculates explicitly the dynamics of a large number of $n\ne 0$ modes. Specifically, we will treat $\hat A=\hat B=\hat{q}^2$, for which
\begin{equation}\label{q2sum}
A\left(\vc{P},\vc{Q}\right)=B\left(\vc{P},\vc{Q}\right)=\sumM Q_n^2 .
\end{equation}

\subsection{System-bath model}
We consider the usual system bath-model,\cite{Tuckerman2010} with classical Hamiltonian
\begin{align}
H =& \frac{p^2}{2m} + V(q) \nonumber\\&+ \sum_{\alpha}\left[\frac{p_\alpha^2}{2\malpha} + \frac{1}{2}\malpha \omegaalpha^2 \left(\xalpha - \frac{\galpha}{\malpha \omegaalpha^2}q\right)^2\right]
\label{eq:system_bath_hamiltonian}
\end{align}
where $V(q)$ is the system potential, $x_\alpha$, $p_\alpha$ and $m_\alpha$ are the positions, momenta and masses of the bath oscillators with frequency $\omegaalpha$ and  coupling coefficients $\galpha$.
To facilitate comparison with exact quantum benchmarks (see below), we calculate the direct-product TCF 
\begin{equation}
    \langle A B(t) \rangle = {1\over Z}\int \ud q\ \ud p\ \ud \vc{p}_{\mathrm{b}}\ \ud \vc{x}\ \ue^{-\beta H_{\mathrm{D}}} A(q,p)\  \ue^{\mathcal{L}t} B(q,p)
\end{equation}
in which the initial distribution is given by the direct-product Hamiltonian
\begin{equation}
    H_{\mathrm{D}} = \frac{p^2}{2m} + V(q) + \sum_{\alpha}\left[\frac{p_\alpha^2}{2\malpha} + \frac{1}{2}\malpha \omegaalpha^2 \xalpha^2\right]
    \label{hathd}
\end{equation}
but the Liouvillian $\mathcal{L}$ is generated from  $H$ of \eqn{eq:system_bath_hamiltonian} and therefore includes the system-bath coupling. 
The equations of motion for $q(t)$ are solved in the same way as in the standard GLE derivation to obtain
\begin{multline}
    m\ddot{q}(t) = -V'[q(t)] - \int_{0}^{t}\ud \tau\sum_\alpha \frac{\galpha^2}{\malpha \omegaalpha^2}\cos\left[\omegaalpha(t-\tau)\right]\dot{q}(\tau) \\
    +\sum_\alpha \galpha\left[\left(\xalpha-\frac{\galpha}{\malpha \omegaalpha^2} q \right)\cos(\omegaalpha t) + \frac{p_\alpha}{\malpha \omegaalpha}\sin(\omegaalpha t)\right].
    \label{eq:full_eom1}
\end{multline}
which can be rewritten as the GLE 
\begin{equation}
m \ddot{q}(t) = -V'[q(t)] - \int_{0}^{t}\ud \tau \zeta(t-\tau) \dot{q}(\tau) + R(t) - q \zeta(t)
\label{eq:gle}
\end{equation}with memory kernel 
\begin{equation}
\zeta(t) = \sum_\alpha \frac{\galpha^2}{\malpha \omegaalpha^2}\cos(\omegaalpha t),
\label{eq:zeta_definition}
\end{equation}
and random force term
\begin{equation}
R(t) = \sum_\alpha \galpha\biggl[\xalpha \cos(\omegaalpha t)
 + \frac{p_\alpha}{\malpha \omegaalpha}\sin(\omegaalpha t)\biggr],
\label{eq:random_force}
\end{equation}
where $\xalpha, p_\alpha$ and $q$ denote the values of these variables at initial time $t=0$. The direct-product GLE of \eqn{eq:gle} is identical to the standard GLE except for the addition of the driving term $-q\zeta(t)$, which acts as a short-term memory of the initial (direct-product) distribution.

It is customary to parametrise the bath via the spectral density 
\begin{equation}
J(\omega) = \frac{\pi}{2} \sum_\alpha \frac{\galpha^2}{\malpha \omegaalpha}\delta(\omega-\omegaalpha),
\end{equation}
which determines both the memory kernel
\begin{equation}
\zeta(t) = \frac{2}{\pi} \int_{0}^{\infty}\ud \omega\ \frac{J(\omega)}{\omega}\cos(\omega t)
\label{eq:zeta_from_J}
\end{equation}
and the random force, which are related by the fluctuation dissipation theorem\cite{Kubo1966}
\begin{equation}
\langle R(0)R(t)\rangle = \frac{1}{\beta}\sum_\alpha \frac{\galpha^2}{\malpha \omegaalpha^2}\cos(\omegaalpha t) = \frac{\zeta(t)}{\beta}.
\end{equation}

Numerical solutions to \eqn{eq:gle} can be propagated using explicit treatment of the bath modes, or equivalently, by discretising
the integrals in \eqnn{eq:gle}{eq:zeta_from_J} to give\cite{Berkowitz1983,Lawrence2019}
\begin{multline}
    - \frac{1}{m} \int_{0}^{t}\ud \tau\ \zeta(t-\tau) p(\tau) \\
    \approx - \frac{\uDelta t}{m}\left(\frac{1}{2} \left[\zeta(0) p_i + \zeta(t_i)p_0 \right] + \sum_{j=1}^{i-1}\zeta(t_j) p_{i-j}\right),
\end{multline}
where $t_j \equiv j\uDelta t$, $q_j \equiv q(t_j)$ and similarly for $p_j$, and 
\begin{equation}
\zeta(t) \approx\frac{2}{\pi} \sum_\alpha w_\alpha \frac{J(\omegaalpha)}{\omegaalpha}\cos(\omegaalpha t),
\label{eq:discretised}
\end{equation}
where $\{w_\alpha\}$ are a set of quadrature weights. Comparing \eqn{eq:discretised} with \eqn{eq:zeta_definition} gives
\begin{equation}
\galpha = \pm\sqrt{\frac{2 w_\alpha J(\omegaalpha)\malpha \omegaalpha}{\pi}}.
\end{equation}
and
\begin{equation}
R(t) = \sum_\alpha g'_\alpha \left[\lambda_\alpha\cos(\omegaalpha t) + \xi_\alpha\sin(\omegaalpha t)\right].
\end{equation}
with
\begin{equation}
g'_\alpha = \galpha \sqrt{\frac{1}{\beta \malpha \omegaalpha^2}} = \sqrt{\frac{2 w_\alpha}{\pi \beta} \frac{J(\omegaalpha)}{\omegaalpha}},
\end{equation}
where $\lambda_\alpha$ and $\xi_\alpha$ are Gaussian random variates with unit variance.

A variety of methods are available for treating the exact quantum version of the system-bath problem. Here, we will use the hierarchical equations of motion (HEOM) approach.\cite{Tanimura1989,Tanimura1991,Tanimura2020,Shi2011,Zhu2012} When applying HEOM, it is simpler (though not essential) to compute the direct-product Kubo TCF
\begin{equation}
    \widetilde{C}_{AB}^\text{D}(t) = \frac{1}{Z_D} \frac{1}{\beta}\int_{0}^{\beta}\ud \lambda\ \mathrm{Tr}\left[\ue^{-(\beta-\lambda)\hat{H}_{\mathrm{D}}}\hat{A}\ue^{-\lambda\hat{H}_{\mathrm{D}}}\ue^{i\hat Ht/\hbar}\hat{B}\ue^{-i\hat Ht/\hbar}\right].
    \label{qdp}
\end{equation}
in which $\hat{H}$ and $\hat{H}_{\mathrm{D}}$  are the quantum Hamiltonians corresponding to the full system-bath $H$ of \eqn{eq:system_bath_hamiltonian} and the direct-product system-bath $H_{\mathrm{D}}$ of \eqn{hathd}, respectively. 
Clearly the neglect of system-bath coupling in $\hat H_D$ means that \eqn{qdp} describes an artificial relaxation process and $\widetilde{C}_{AB}^\text{D}(t)$ can be expected to be different to $\widetilde{C}_{AB}(t)$ (unless the bath is weak). This artificiality does not prevent us from using \eqn{qdp} to benchmark Matsubara dynamics; in fact, it is advantageous, since the thermalisation depends critically on the relaxation dynamics of the non-centroid Matsubara modes  $(P_n,Q_n)$, making comparison with \eqn{qdp} a stringent test.

\section{System-bath Matsubara Dynamics}

In this Section, we derive the Matsubara approximation to the direct-product Kubo TCF of \eqn{qdp} (Secs.~IIIA and IIIB), and develop a real-noise approximation to it (Sec. IIIC). These derivations can easily be modified to give the Matsubara approximation to the equilibrium Kubo TCF of \eqn{qexact} (see Appendix~\ref{app:equilibrium}).

\subsection{Matsubara GLE}\label{sec:matsubara_gle}

The system-bath Matsubara Hamiltonian corresponding to $H$ of \eqn{eq:system_bath_hamiltonian} is easily shown to be
\begin{multline}
    H_M\left(\vc{Q}, \vc{P}, \vc{X}, \vc{P}_{\mathrm{b}} \right) = \sum_{n} \frac{P_n^2}{2m} + U_M(\vc{Q})\\
    + \sum_{\alpha,n}\left[\frac{P_{\alpha n}^2}{2\malpha} + \frac{\malpha \omegaalpha^2}{2}\left(X_{\alpha n} -\frac{\galpha Q_n}{\malpha \omegaalpha^2}\right)^2\right],
    \label{eq:ham_mats_sys_bath}
\end{multline}
where the potential of mean force $U_M(\vc{Q})$ is obtained by inserting the system potential $V(q)$ into \eqn{eq:MF_W} of Appendix~\ref{app:mean_force}. Note that each Matsubara mode $Q_n$ is coupled to its own bath $\{X_{\alpha n}\}$ and $\{P_{\alpha n}\}$, and that the only term that couples Matsubara modes with different $n$ is the system potential $ U_M(\vc{Q})$. Since we wish to approximate $\widetilde{C}^\text{D}(t)$ of \eqn{qdp}, we also need the Matsubara Hamiltonian
corresponding to ${H}_{\mathrm{D}}$ of \eqn{hathd}, which is 
\begin{align}
H_M^\text{D}\left(\vc{Q}, \vc{P}, \vc{X}, \vc{P}_{\mathrm{b}}\right) =& \sum_{n} \frac{P_n^2}{2m} + U_M(\vc{Q})
\nonumber\\
&+ \sum_{\alpha,n}\left[\frac{P_{\alpha n}^2}{2\malpha} + \frac{\malpha \omegaalpha^2}{2}X_{\alpha n}^2\right]
\label{eq:dp_hamiltonian}
\end{align}
The Matsubara approximation to the direct-product Kubo TCF of \eqn{qdp} can then be written
\begin{multline}
    \widetilde{C}_{AB}^{\ \mathrm{D}[M]}(t) = \frac{1}{Z}\int\ud\, \vc{Q}\,\ud \vc{P}\, \ud \vc{X}\, \ud \vc{P}_{\mathrm{b}}\\
    \ue^{-\beta H_M^\text{D}\left(\vc{Q}, \vc{P}, \vc{X}, \vc{P}_{\mathrm{b}} \right)}\ue^{\ui\beta \theta\left(\vc{Q}, \vc{P}, \vc{X}, \vc{P}_{\mathrm{b}} \right)} A\left(\vc{Q}\right)B\left[\vc{Q}(t)\right],
    \label{eq:matsubara_tcf_direct_product}
\end{multline}
where \(\int\ud \vc{X} = \prod_{\alpha,n} \int\ud X_{\alpha n} \) (and similarly for $\vc{P}_{\mathrm{b}}$),  and 
\begin{align}
    \theta\left(\vc{Q}, \vc{P}, \vc{X}, \vc{P}_{\mathrm{b}}\right) =& \underbrace{\sum_{n} \omegan P_n Q_{\mybar{n}}}_{\theta_\mathrm{S}} + \underbrace{\sum_{n,\alpha} \omegan P_{\alpha n} X_{\alpha \mybar{n}}}_{\theta_\mathrm{B}}
    \label{eq:sysbath_phase}
\end{align}
is the Matsubara phase. By analogy with the standard derivatgion of a classical GLE,\cite{Tuckerman2010} it is
straightforward to show  that 
\begin{align}
    m\ddot{Q}_n(t) =& -\frac{\partial U_M (\vc{Q}(t))}{\partial Q_n}\nonumber\\
    &- \int_{0}^{t}\ud \tau\sum_\alpha \frac{\galpha^2}{\malpha \omegaalpha^2}\cos\left[\omegaalpha(t-\tau)\right]\dot{Q}_n(\tau) \nonumber\\
    &+ G(\vc{Q}, \vc{X}, \vc{P}_{\mathrm{b}})- Q_n\underbrace{\sum_\alpha \frac{\galpha^2}{\malpha \omegaalpha^2}\cos(\omegaalpha t)}_{\zeta(t)},
    \label{eq:mats_gle_expanded}
\end{align}
where
\begin{align}
    G(\vc{Q}, \vc{X}, \vc{P}_{\mathrm{b}})=& \sum_\alpha \galpha\left[X_{\alpha n}\cos(\omegaalpha t)\right.
    \nonumber\\
    &+ \left.\frac{P_{\alpha n}}{\malpha \omegaalpha}\sin(\omegaalpha t)\right].
\end{align}
The Matsubara equations of motion \eqn{eq:mats_gle_expanded} resemble a generalisation to $M$ modes of the classical \eqn{eq:gle}. \footnote{It should be noted, however, that a different driving term of purely quantum origin is obtained for the equilibrium initial condition (see Appendix~\ref{app:equilibrium})}  However, $G$ does not have the form of a random force term with Gaussian noise, because of the bath phase $\theta_\mathrm{B}$ in \eqn{eq:sysbath_phase}. We therefore remove this phase (without making any approximation\footnote{It should be noted that apart from making the substitution, one must move the limits of the integrals in \eqn{eq:matsubara_tcf_direct_product} back onto the real line. This is justified by Cauchy's residue theorem only if the integrand has no poles between the real line and the ${P}=-\ui m \omegan Q_{\mybar{n}}$. This can be guaranteed at $t=0$ and is likely even at other times. A presence of a pole would lead to unstable trajectories, but the presence of unstable trajectories is not in itself a proof of the presence of a pole. However, this is of little concern, since this computational approach should not be used for any non-negligible fraction of unstable trajectories as shown in the numerical results section.}) by analytically continuing the bath modes, which amounts to replacing each ${P}_{\alpha n}$ by $P_{\alpha n} + \ui \malpha \omegan X_{\alpha \mybar{n}}$. \Eqn{eq:matsubara_tcf_direct_product} then becomes
 \begin{multline}
    \widetilde{C}_{AB}^{\ \mathrm{D}[M]}(t) = \frac{1}{Z}\int\ud \vc{Q}\ \ud \vc{P}\ \ud \vc{X}\ \ud \vc{P}_{\mathrm{b}}\\
    \ue^{-\beta R_M^\text{D}\left(\vc{Q}, \vc{P}, \vc{X}, \vc{P}_{\mathrm{b}} \right)}\ue^{\ui\beta \theta_\text{S}\left(\vc{Q}, \vc{P} \right)} A\left(\vc{Q}\right)B\left[\vc{Q}(t)\right].
    \label{notb}
\end{multline}
where
 \begin{multline}
R_M^\text{D}\left(\vc{Q}, \vc{P}, \vc{X}, \vc{P}_{\mathrm{b}}\right) =\\
 \sum_{n} \frac{P_n^2}{2m} + U_M(\vc{Q})
+ \sum_{\alpha,n}\left[\frac{{P}_{\alpha n}^{2}}{2\malpha} + \frac{\malpha \omega_{\alpha n}^2}{2}X_{\alpha n}^2\right] ,
\label{eq:dp_distribution}
\end{multline}
with $\omega_{\alpha n}^2 = \omegaalpha^2 + \omegan^2$, and the equations of motion retain the form of \eqn{eq:mats_gle_expanded} except that $G(\vc{Q}, \vc{X}, \vc{P}_{\mathrm{b}})$ is replaced by 
\begin{multline}
G^{\mathbb{C}}(\vc{Q}, \vc{X}, \vc{P}_{\mathrm{b}})=\sum_\alpha \galpha \left[X_{\alpha n}\cos(\omegaalpha t)\right.\\
+ \left.\left(\frac{{P}_{\alpha n}}{\malpha \omegaalpha} + \ui  \frac{\omegan}{\omegaalpha} X_{\alpha \mybar{n}}\right) \sin(\omegaalpha t)\right].
\label{eq:noise_driving_dp}
\end{multline}
The variables $\vc{P}_{\mathrm{b}}$ and $\vc{X}$ have Gaussian distributions in \eqn{notb}, and hence we can now write the equations of motion in the form of a GLE,
\begin{equation}
\begin{split}
m\ddot{Q}_n(t) = &-\frac{\partial U_M(\vc{Q})}{\partial Q_n} - \int_{0}^{t}\ud \tau\ \zeta(t-\tau)\dot{Q}_n(\tau)\\
& + R_n^{\mathbb{C}}(t) - Q_n\zeta(t).
\end{split}\label{eq:matsubara_gle_dp}
\end{equation}
with a {\em complex} random force
\begin{align}
R_n^{\mathbb{C}}(t) =& \sum_\alpha g'_\alpha \left[\frac{\omegaalpha}{\omega_{\alpha n}}\lambda_{\alpha n}\cos(\omegaalpha t)\right.\nonumber\\
&+ \left.\left(\xi_{\alpha n} + \ui  \frac{\omegan}{\omega_{\alpha n}} \lambda_{\alpha \mybar{n}}\right) \sin(\omegaalpha t)\right]
\label{eq:noise_complex_sampling}
\end{align}
where $\lambda_{\alpha n}$ and $\xi_{\alpha n}$ are Gaussian variates with unit variance. (Note that the random numbers in the imaginary parts of the sine coefficients are the same as those in the cosine coefficients of opposite $n$.)

We will refer to \eqn{eq:matsubara_gle_dp} as the {\em Matsubara GLE}. This equation looks superficially like a simple generalisation to Matsubara modes of the classical GLE (\eqn{eq:gle}), and it reduces to CMD (classical dynamics on the centroid potential of mean force) when $M=1$.
 However, when $M>1$, the complex noise generated by \eqn{eq:noise_complex_sampling} makes a major difference, since it satisfies a set of quantum fluctuation-dissipation relations
 \begin{subequations}\label{both}
\begin{align}
    \langle R_n^{\mathbb{C}}(t_1) R_{n}^{\mathbb{C}}(t_2) \rangle =& \frac{\zeta(t_2-t_1) - K_n (t_2-t_1)}{\beta} \\
      \langle R_n^{\mathbb{C}}(t_1) R_{\mybar{n}}^{\mathbb{C}}(t_2) \rangle=& -\ui\frac{L_n (t_2-t_1)}{\beta}\label{good}\end{align}
       \end{subequations}
where 
 \begin{subequations}
\begin{align}
   K_n (t)=&\frac{2\omegan^2}{\pi} \int_{0}^{\infty}\ud \omega \frac{J(\omega)}{\omega} \frac{\cos(\omega t)}{\omega^2 + \omegan^2}\label{eq:K}\\
   L_n (t)=&\frac{2\omegan}{\pi}\int_{0}^{\infty}\ud\omega\ J(\omega)\frac{\sin(\omega t)}{\omega^2 + \omegan^2}.\label{eq:L}
\end{align}
       \end{subequations}
These relations ensure that the bath acts as a quantum thermostat such that the system Matsubara modes equilibrate  to the (exact) quantum Boltzmann distribution. We emphasise that the only approximation made to obtain \eqn{eq:matsubara_gle_dp} from the exact quantum dynamics is to have smoothed the imaginary-time Feynman paths. \Eqn{eq:matsubara_gle_dp} is therefore exact, except for the neglect of real-time coherence. 


\subsection{Analytically continuing the system variables}
The presence of the system Matsubara phase $\theta_\mathrm{S}$ in \eqn{notb} makes the integral very hard to integrate numerically (except when $M=1$ which gives CMD, with $\theta_\mathrm{S}=0$). We therefore convert $\theta_\mathrm{S}$ to ring-polymer springs, by analytically continuing $P_n$, as was done for the bath modes in Sec~\ref{sec:matsubara_gle}. This is equivalent to replacing each $P_n$ in \eqn{notb} by $P_n + \ui m \omegan Q_{\mybar{n}}$, so that it becomes
\begin{multline}
    \widetilde{C}_{AB}^{\ [M]}(t) = \frac{1}{Z}\int\ud \vc{Q}\ \ud \vc{P}\ \ud \vc{X}\ \ud \vc{P}_{\mathrm{b}}\\
    \ue^{-\beta S_M^\text{D}\left(\vc{Q}, \vc{P}, \vc{X}, \vc{P}_{\mathrm{b}} \right)} A\left(\vc{Q}\right)B\left[\vc{Q}(t)\right]
    \label{noph}
\end{multline}
where
\begin{multline}
S_M^\text{D}\left(\vc{Q}, \vc{P}, \vc{X}, \vc{P}_{\mathrm{b}}\right) =
 \sum_{n} \left[\frac{P_n^2}{2m}+ \frac{1}{2}m\omegan^2 Q_n^2\right] + U_M(\vc{Q})\\
+ \sum_{\alpha,n}\left[\frac{{P}_{\alpha n}^{2}}{2\malpha} + \frac{\malpha \omega_{\alpha n}^2}{2}X_{\alpha n}^2\right] 
\label{eq:dp_distribution}
\end{multline}
and the equations of motion become\cite{Hele2015a}
\begin{align}
    \begin{split}
        \dot{{P}}_n &= -\frac{\partial U_M(\vc{Q}(t))}{\partial Q_n} - m\omegan^2 Q_n - \ui \omegan {P}_{\mybar{n}}\\
        & \qquad- \frac{1}{m} \int_{0}^{t}\ud \tau\ \zeta(t-\tau) \left[{P}_n(\tau) + \ui m \omegan Q_{\mybar{n}}(\tau)\right]
        \\
        & \qquad  + R_n^{\mathbb{C}}(t) - Q_n(0)\zeta(t)
    \end{split}\nonumber
    \\
    \dot{Q}_n &= \frac{{P}_n}{m} + \ui \omegan Q_{\mybar{n}}.\label{eq:ac_eom_q}
\end{align}
We emphasise that no additional approximation has been made here: \eqnt{noph}{eq:ac_eom_q} are equivalent to \eqnt{notb}{eq:noise_complex_sampling}.

When the bathless versions of \eqnt{noph}{eq:ac_eom_q} were derived in ref.~\onlinecite{Hele2015a}, it was pointed out that, although the analytic continuation has eliminated the phase, it has not eliminated the numerical difficulty, because the dynamics now takes place in the complex plane, and is found to be pathologically unstable (unless $V(q)$ is harmonic, $M=1$, or the imaginary parts of the equations of motion are discarded to give RPMD). However, a surprising result reported in Sec.~IV below  is that the bath  {\em stabilises} the dynamics, up to some maximum value $M$ which depends on the bath strength and the anharmonicity of $V$. 

\subsection{Real-noise approximation}\label{sec:decorrelation}

One might expect that the stabilization just mentioned would be better if the random-force term were real, since the imaginary part of this term will tend to kick the trajectories further along the imaginary axis. We therefore propose approximating the Matsubara GLE 
 by
\begin{equation}
    \begin{split}
    m\ddot{Q}_n(t) = &-\frac{\partial U_M(\vc{Q}(t))}{\partial Q_n}  - \int_{0}^{t}\ud \tau\ \zeta(t-\tau)\dot{Q}_n(\tau)\\
    &\qquad + R_n(t) - Q_n(0)\zeta(t)
    \end{split}
\label{really}
\end{equation}
which is identical to \eqn{eq:matsubara_gle_dp}, except that the imaginary part has been discarded from $R_n^{\mathbb{C}}(t)$, leaving the real part
\begin{equation}
R_n(t) = \sum_\alpha  \frac{g'_\alpha \omegaalpha}{\omega_{\alpha n}}\left[\lambda_{\alpha n}\cos(\omegaalpha t)
+ \xi_{\alpha n} \sin(\omegaalpha t)\right].
\label{eq:noise_real_sampling}
\end{equation}
Equations~(\ref{noph})--(\ref{eq:ac_eom_q}) remain the same, except that $R_n^{\mathbb{C}}(t)$ is replaced by 
$R_n(t)$ in \eqn{eq:ac_eom_q}.

To gauge how well this approximation is likely to work, we mention one point in its favour, and one against. 
First, it was noted in refs.~\onlinecite{Ivanov2010} and \onlinecite{Benson2021} that one can eliminate the Matsubara phase by decorrelating the initial distribution in the TCF, instead of by analytic continuation. When applied to the initial distribution of the bath modes
            \begin{multline}
        \Phi\left(\vc{X}, \vc{P}_{\mathrm{b}} \right)= \prod_{\alpha,n}\exp \biggl(-\beta\biggl[
        \frac{P_{\alpha n}^2}{2\malpha} \\
        + \frac{\malpha \omegaalpha^2}{2}X_{\alpha n}^2 - \ui \omegan P_{\alpha n} X_{\alpha \mybar{n}}\biggr]
        \biggr)
\end{multline}
in \eqn{eq:matsubara_tcf_direct_product}, this `decorrelation approximation' gives
\begin{multline}
\Phi \left(\vc{X}, \vc{P}_{\mathrm{b}}; \vc{Q}\right)\simeq\\
\frac{\int \Phi_{\mathrm{b}}\left(\vc{X}, \vc{P}_{\mathrm{b}}; \vc{Q}\right) \ud \vc{X} \times \int \Phi_{\mathrm{b}}\left(\vc{X}, \vc{P}_{\mathrm{b}}; \vc{Q}\right) \ud \vc{P}_{\mathrm{b}}}
{\int \Phi_{\mathrm{b}}\left(\vc{X}, \vc{P}_{\mathrm{b}}; \vc{Q}\right) \ud \vc{X} \ud \vc{P}_{\mathrm{b}}}\\
= \prod_{\alpha,n}{\beta \omegaalphan^2\over2\pi \omegaalpha}\exp \biggl(-\beta \biggl[
\frac{P_{\alpha n}^2 \omegaalphan^2}{2\malpha \omegaalpha^2} + \frac{\malpha \omegaalphan^2}{2} X_{\alpha n}^2
\biggr]\biggr)\label{messy}
\end{multline}
If we then retrace the steps in Secs.~IIIA and B (minus the analytic continuation of the bath modes), we obtain \eqnn{really}{eq:noise_real_sampling}. Hence, the neglect of the imaginary component of the noise is equivalent to approximating the initial distribution (by decorrelating the bath modes as in \eqn{messy}), but it makes no approximation to the Matsubara dynamics equations of motion. We might therefore expect the errors due to neglecting $\text{Im}[R_n^{\mathbb{C}}(t)]$ to be minor, since the neglect of correlations between the bath-modes will only take effect when bath modes with different $n$ have had sufficient time to interact via the system potential $V$. However, the use of \eqn{messy} also shows that the neglect of $\text{Im}[R_n^{\mathbb{C}}(t)]$ modifies the quantum fluctuation-dissipation relations of \eqn{both} to
        \begin{subequations}\label{hancock}
\begin{align}\label{apfd}
    \langle R_n(t_1) R_{n}(t_2) \rangle = & \ \frac{\zeta(t_2-t_1) - [K_n (t_2-t_1)+ K_n (t_2+t_1)]/2}{\beta} \\
      \langle R_n(t_1) R_{\mybar{n}}(t_2) \rangle= & \ 0\label{bad}\end{align}
             \end{subequations}
In other words, the Matsubara dynamics of the bath is insufficiently ergodic\footnote{Presumably because the dynamics conserves the Matsubara phase.} for it to recover completely from the initial decorrelation of $X_{\alpha n}$ and $P_{\alpha \mybar{n}}$, which prevents it from thermostatting correctly. We can therefore expect  the real-noise approximation to the TCF not to tend to the exact $t\to\infty$ limit.
In Sec.~IV, we find numerically that these errors decrease with increase in bath strength and  the stability gained by neglecting $\text{Im}[R_n^{\mathbb{C}}(t)]$ is considerable. 

\begin{figure}
    \includegraphics{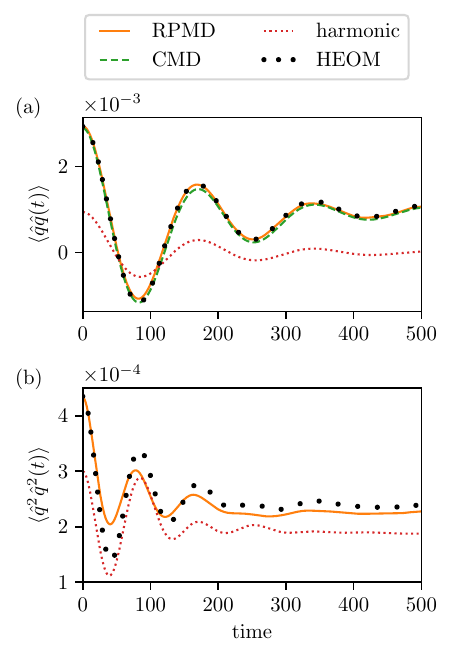}
    \caption{Time-correlation functions (TCFs) computed using standard methods for a Morse oscillator coupled to a harmonic bath, with the parameters of Table~\ref{tab:1}. All TCFs were initiated with direct-product initial conditions (see \eqn{qdp}). All units are in a.u.} 
    \label{fig:morse}
\end{figure}

\section{Numerical tests of the Matsubara GLE}\label{sec:results}

\begin{table}
\begin{tabular}{ l l l }
 System  & $D_0$ & $D_{0,\mathrm{water}}/2$ \\
  & $D_{0,\mathrm{water}}$ & 0.18748  \\
     &$\omega_{\mathrm{well}}$&$1.70304\times 10^{-2}$\\
     & $a$ &  $\omega_{\mathrm{well}}\sqrt{m/(2D_0)}$ \\
   & $m$ & 1741.1 \\
    & & \\
 Bath  &$\eta$&$2\eta_{\mathrm{crit}}=2\times2m\omega_{\mathrm{well}}$ \\
   &$\omegac$&$\omega_{\mathrm{well}}$\\
      &&\\
   Temperature\ \ \ & T & 150 K
\end{tabular}
\caption[Table caption text]{Parameters (in a.u., unless indicated otherwise) used in the coupled oscillator-bath model of \eqnn{morse}{debye} to calculate the results in Figs.~\ref{fig:morse}, \ref{fig:noisy} and \ref{fig:convergence_correction}. In Fig.~\ref{fig:stability}, when $D_0$ is changed, $a$ is changed accordingly such that $\omega_{\mathrm{well}}$ stays constant. $\eta_{\mathrm{crit}}$ is the value of the coupling constant that would give critical damping for a classical harmonic oscillator in the $\omega_{\mathrm{c}}\to\infty$ limit.}
\label{tab:1}
\end{table}

In this Section we test the stability of the analytically continued system Matsubara dynamics, for both complex and real noise,  then investigate how far this takes us towards a numerically converged Matsubara-dynamics approximation to
the direct-product Kubo TCF $\langle \hat q^2\hat q^2(t)\rangle$. 

All calculations used a Morse system potential
\begin{equation}\label{morse}
    V(q) = D_0 \left(1-\mathrm{e}^{-a q}\right)^2
\end{equation}
and a Debye bath spectral density
\begin{equation}\label{debye}
    J(\omega) = \eta \omega \frac{\omegac ^2}{\omegac ^2+\omega^2},
\end{equation}
for which expressions for all the bath properties (such as $g_{\alpha n}$, $K_n$, $L_n$, etc.) were derived (and are given in Appendix~\ref{app:baths}).

Figure~\ref{fig:morse} shows direct-product Kubo TCFs  $\langle \hat q\hat q(t)\rangle$ and $\langle \hat q^2\hat q^2(t)\rangle$, computed using the parameters of Table~\ref{tab:1}, for a variety of standard methods. Evidently, this choice of parameters makes the dynamics significantly anharmonic. However,  the anharmonicity causes only very weak coupling of the dynamics of the centroid to the dynamics of the $n\ne0$ Matsubara `fluctuation'  modes, since the RPMD and CMD linear $\langle \hat q\hat q(t)\rangle$ are in close agreement with HEOM, indicating that each of these methods gives a good description of the dynamics of the system centroid as it relaxes from the direct-product initial condition into equilibrium with the bath. These findings are consistent with previous applications of non-equilibrium RPMD to system-bath problems.\cite{Welsch2016}

However, the TCF we will focus on in what follows is the non-linear $\langle \hat q^2\hat q^2(t)\rangle$. This TCF depends explicitly on the dynamics of the $n\ne0$ fluctuation modes, through \eqn{q2sum}, and it is well known\cite{Craig2004,Althorpe2021} that RPMD and CMD do not describe this dynamics correctly. Figure~\ref{fig:morse} shows that the RPMD $\langle \hat q^2\hat q^2(t)\rangle$ agrees with HEOM at $t=0$ and at longer times, as expected (since it gives the correct quantum Boltzmann statistics in these limits), but is otherwise in poor agreement. The CMD results for $\langle \hat q^2\hat q^2(t)\rangle$ are in extremely poor agreement with HEOM (and are not shown), for the obvious reason that CMD follows only the dynamics of the centroid, whereas most of the contribution to $\langle \hat q^2\hat q^2(t)\rangle$ comes from the $n\ne0$ fluctuation modes. Our goal in what follows is to test numerically whether Matsubara dynamics is capable of describing the dynamics of these modes and thus giving good agreement with HEOM for $\langle \hat q^2\hat q^2(t)\rangle$.

\begin{figure}
    \includegraphics[scale=1]{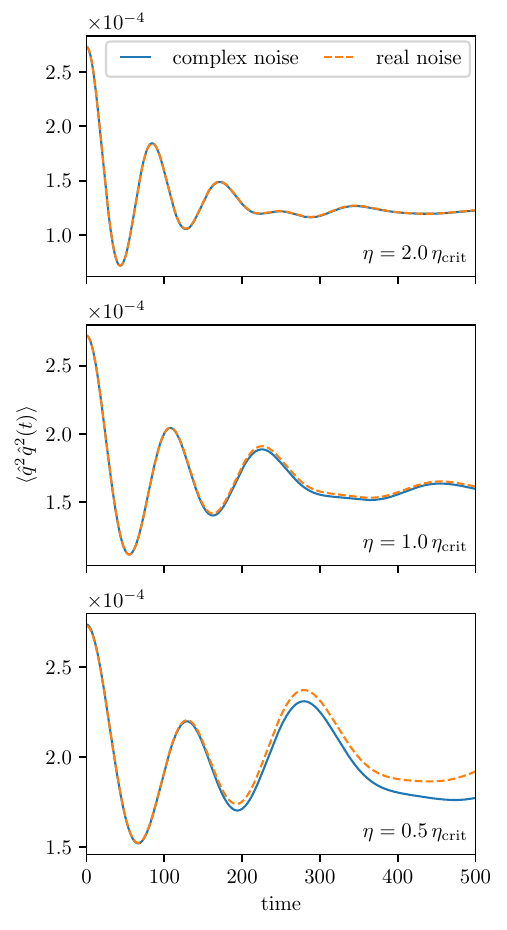}
    \caption{Matsubara dynamics calculations of the $\langle \hat{q}^2 \hat{q}^2(t)\rangle$ TCF of Fig.~1(b), for $M=25$ Matsubara modes, obtained by solving the Matsubara GLE \eqn{eq:matsubara_gle_dp} (complex noise) and its real-noise approximation \eqn{really} (real noise). All units are in a.u.}
    \label{fig:noisy}
\end{figure}

\subsection{Matsubara GLE with complex noise}
As a first step, we tested the Matsubara GLE of \eqn{eq:matsubara_gle_dp} using the parameters of Table~\ref{tab:1}. The system was equilibrated in a standard (one-dimensional) T(RPMD) simulation with $N=256$ polymer beads, from which analytically continued Matsubara GLE slave trajectories were propagated until $t= 500$ a.u.\ using the algorithm of Appendix~\ref{propagator} (with a timestep of 0.1~a.u.\ and including 1000 implicit bath modes). The Matsubara potential of mean force $U_M$ was generated on the fly, following the procedure
of ref.~\onlinecite{Trenins2018}, in which ring-polymer springs are attached to the $N-M$ highest frequency ring-polymer normal modes, which follow adiabatically separated dynamics\cite{Cao1994c,Hone2006} subject to a strong PILE thermostat; an adiabaticity parameter of $\Gamma=16$ was sufficient to converge the TCFs. The discretisation of the bath frequencies was done using the efficient approach of ref.~\onlinecite{Craig2007}.

Surprisingly, we found that the coupling to the bath made the analytically continued system dynamics numerically stable for $M\le45$. We also found that the number of trajectories needed to converge the TCFs was reasonable, with $2\times10^5$ trajectories being more than sufficient to converge the TCFs to graphical accuracy (and $2\times10^4$ trajectories and just 100 implicit bath modes being  sufficient to converge to within 1\%).  The resulting  $\langle \hat q^2\hat q^2(t)\rangle$ TCF for $M=25$ is shown in Fig.~\ref{fig:noisy}. While the complex noise dynamics can be pushed up to $M=45$, the resulting $\langle \hat q^2\hat q^2(t)\rangle$ TCF (not shown) still falls short of the exact quantum result shown in Fig.~\ref{fig:morse}, indicating that many more Matsubara modes would be required for numerical convergence.

\begin{figure}
    \includegraphics{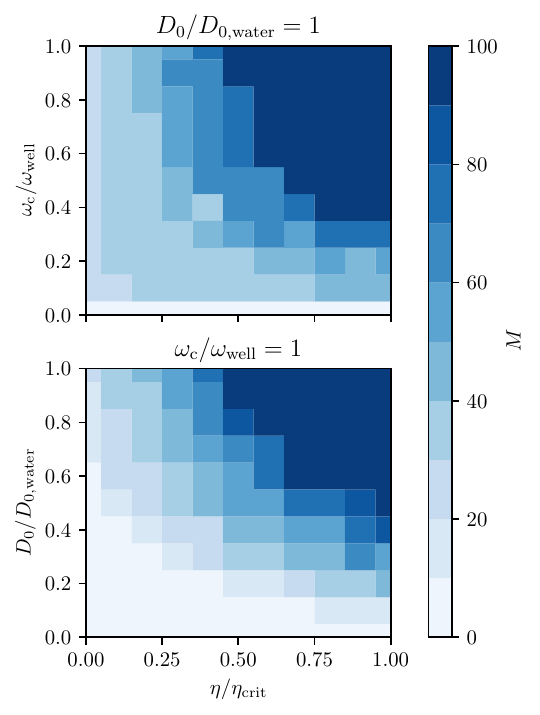}
    \caption{Plots showing the largest number of Matsubara modes $M$ at which the analytically continued trajectories are numerically stable, as a function of the system anharmonicity (which increases with decreasing $D_0$) and the bath strength (characterised by the Debye bath parameters $\eta$ and $\omegac$). Trajectories are defined to be numerically stable when fewer than 0.1~\% of them diverge (see Sec.~IVB).}
    \label{fig:stability}
\end{figure}

\subsection{Matsubara GLE with real noise}

We then repeated the calculation using the real-noise approximation of \eqnn{really}{eq:noise_real_sampling}, keeping all other details of the calculation the same as Sec.~IVA. As discussed in Sec.~IIIC, the real-noise approximation is not guaranteed to thermalise correctly, because it erroneously replaces the fluctuation-dissipation relations of \eqn{both} by \eqn{hancock}. However, Fig.~\ref{fig:noisy} shows that the resulting thermostatting error is small, and that it decreases with increase in $\eta$, such that  $\eta=2\eta_\text{crit}$ gives errors which are within graphical accuracy for $M=25$ and the error for $\eta=\eta_\text{crit}$ is still very small (within 1.6~\%). Thus the  imaginary component of the noise has only a subtle effect on the equilibration dynamics once $\eta$ exceeds a certain amount (which in this system is about $\eta=\eta_\text{crit}$); future work will be needed to determine why this is so. 

Very usefully, the real-noise approximation was found to make the analytically continued Matsubara dynamics much more stable, increasing the maximum value of $M$ from $M=45$ (complex noise) to $M\approx200$ (real noise). Since the stability is likely to depend on the bath strength and the anharmonicity of $V(q)$, we varied the bath parameters $\eta$ and $\omegac$, and the system parameter $D_0$, to investigate the maximum value of $M$ at which fewer than
 $0.1\%$ of the trajectories diverged. 
This criterion was used because it was found to ensure that the gaps between TCFs computed using $M$ and $M+2$ modes decreased on increasing $M$.\footnote{This is the expected convergence behaviour, since the quantum Boltzmann weight can be expected to decreases monotonically for sufficiently large $M$. This trend is not obeyed once the trajectories become unstable, causing the integrals to diverge.}  Figure~\ref{fig:stability} shows that, as expected, stability increases with increasing bath strength (i.e.\ increasing either $\eta$ or $\omegac$), and decreases with increasing anharmonicity in $V(q)$. Note that, even for $\eta<\eta_\text{crit}$, a considerable number of Matsubara modes can still be propagated stably.

\begin{figure}
    \includegraphics{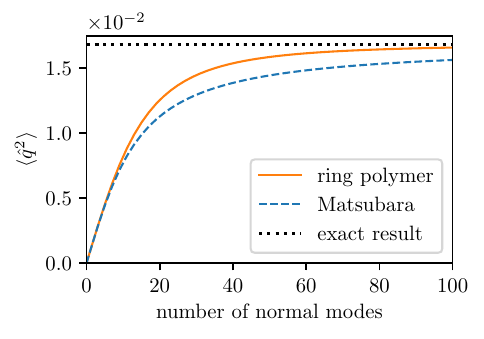}
    \caption{Illustration of the `Matsubara tail' for $\langle \hat{q}^2 \rangle$ for a harmonic oscillator of frequency $\omega_\text{well}$ at 150 K. Note the much slower convergence with respect to the number of Matsubara modes than the number of ring-polymer normal modes.  }
    \label{fig:rp_mats_convergence}
\end{figure}

\subsection{Harmonic tail correction}

Although $M=200$ would appear to be a large number of Matsubara modes, the $\langle\hat q^2\hat q^2(t)\rangle$ TCF requires a much larger value of $M$ for numerical convergence. This is on account of the slowness with which the 
the explicit sum over the Matsubara modes in \eqn{q2sum} converges with respect to $M$.\cite{Ceperley1995,Coalson1986,Doll1999,Chakravarty1998} We can gauge the likely length of this `Matsubara tail' by computing the (static) thermal expectation value of $\hat q^2$ at 150 K  for a (bathless) harmonic oscillator with frequency $\omega_\text{well}$. Figure~\ref{fig:rp_mats_convergence} shows that thousands of Matsubara modes are needed to converge to a result attainable with just 100 ring-polymer beads. This is the price one pays for smoothing the imaginary-time Feynman paths. 

However, for sufficiently high $n$,  we expect the Matsubara dynamics to become harmonic, on account of the tightness of the ring-polymer springs (or equivalently the highly oscillatory nature of the Matsubara phase), which confines $Q_n$ to a small region around  zero. This  is borne out by Fig.~\ref{fig:9cases} of Appendix~\ref{app:harmonic_correction}, which shows that the centroid dynamics is much more anharmonic than that of the $n\ne 0$ fluctuation modes, which quickly become harmonic with increasing $n$.\footnote{This finding is consistent with the assumptions of the planetary model of refs.~\onlinecite{Smith2015} and \onlinecite{Willatt2018}.}  In the harmonic limit, the Matsubara GLE splits into pairs of GLEs, each of which couples only the modes $n$ and $\mybar{n}$. These equations are straightforward to solve numerically, since analytically continued harmonic
trajectories are stable. 

\begin{figure}
    \includegraphics{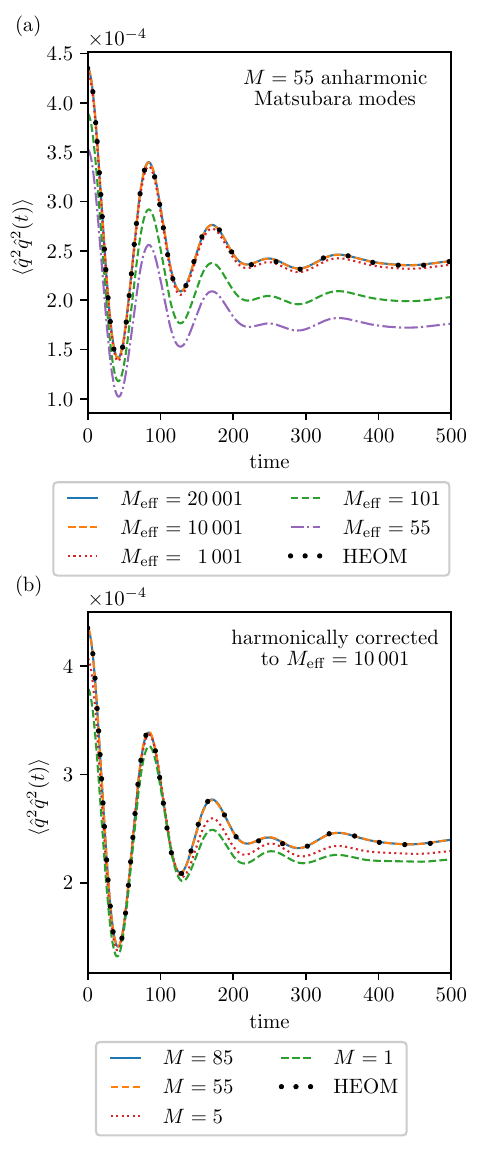}
    \caption{Convergence of the real-noise Matsubara TCF $\langle \hat{q}^2 \hat{q}^2(t)\rangle$ (using the parameters of Table~\ref{tab:1}) with respect to (a) the total number of Matsubara modes $M_{\mathrm{eff}}$ and (b) the number of non-harmonically treated Matsubara modes $M$. The HEOM results are also shown. All units are in a.u.}
    \label{fig:convergence_correction}
\end{figure}

We therefore solved the analytically continued (real-noise) Matsubara GLE (\eqn{really}) for $M$ modes,  then generated harmonic corrections for the $|n|>(M-1)/2$ modes up to a large value
$|n|=(M_\text{eff}-1)/2$  (as described in Appendix~\ref{app:harmonic_correction}). For the system of Table~\ref{tab:1}, this approach converged the $\langle \hat q^2\hat q^2(t)\rangle$ TCF  using $M=55$ and $M_\text{eff}=10\,001$, as demonstrated in Fig.~\ref{fig:convergence_correction} (where identical curves are obtained to within graphical accuracy for $M=55$ and $M=85$, demonstrating that the dynamics of the $|n|>(M-1)/2$ Matsubara modes is indeed harmonic).

\subsection{Agreement between Matsubara dynamics and HEOM}

Figure~\ref{fig:convergence_correction} shows that the $\langle \hat q^2\hat q^2(t)\rangle$ TCF computed using the real-noise Matsubara GLE for $\eta=2\eta_\text{crit}$ is in almost perfect agreement with the numerically exact HEOM results.  We also found that the linear $\langle \hat q\hat q(t)\rangle$ real-noise Matsubara TCF  (not shown) agrees with the HEOM results to within graphical accuracy, and has converged by $M=15$ (thus correcting the very small discrepency between CMD and HEOM in Fig.~\ref{fig:morse}).

\begin{figure}
    \includegraphics{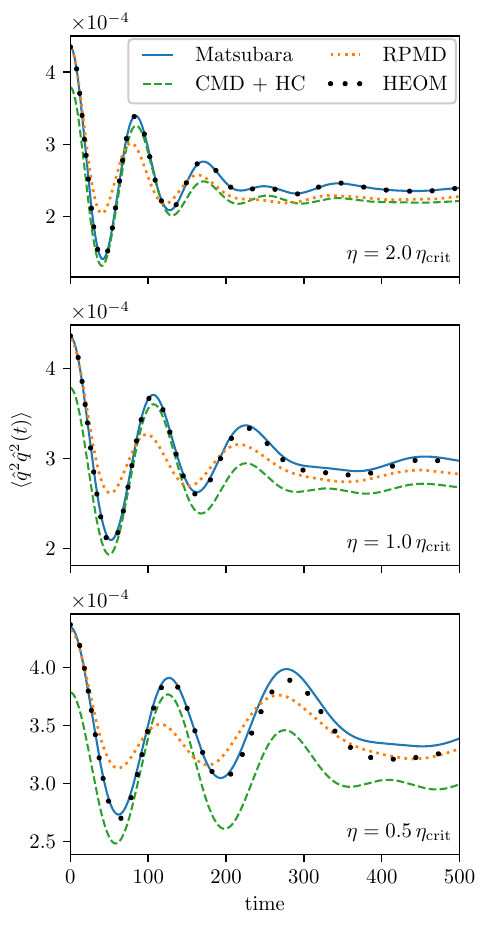}
    \caption{Real-Noise Matsubara TCFs $\langle \hat{q}^2 \hat{q}^2(t)\rangle$, compared with HEOM, RPMD and harmonically corrected CMD results, for three different bath strengths. The Matsubara calculations used $M=55, 55, 35$ for $\eta/\eta_{\mathrm{crit}} = 2, 1, 0.5$, with $M_{\mathrm{eff}} = 10\,001$. Harmonically corrected CMD is equivalent to Matsubara dynamics with $M=1$ and  $M_{\mathrm{eff}} = 10\,001$. All units are in a.u.}
    \label{fig:3cases}
\end{figure}

Figure~\ref{fig:3cases} shows what happens when we apply the Matsubara GLE at weaker bath-strengths. For
$\eta=\eta_\text{crit}$ the Matsubara TCF is in very close agreement with the HEOM result, with a barely noticeable drift at longer times. For
$\eta=0.5\eta_\text{crit}$, the Matsubara and HEOM TCFs are also in close agreement until about $t=250$~a.u., after which the errors from the real-noise approximation (cf~Fig.~2) cause a significant drift in the thermalisation. 

Finally, we also plot in Fig.~6 (see also Fig.~5b) the result of truncating the calculation at $M=1$, with all modes $0<|n|<(M_\text{eff}-1)/2$ included using the harmonic correction of Sec.~IVC. This approach is equivalent to using CMD to describe the motion of the centroid and treating the fluctuation dynamics harmonically. These results still have significant errors, but they are a marked improvent on the purely harmonic result (see Fig.~1 for $\eta=2\eta_\text{crit}$); this is consistent with the finding of Sec.~IVC that the dynamics of the centroid is much more anharmonic than that of the fluctuations. 

\section{Conclusions}

The results of Sec.~IV are the first converged Matsubara dynamics TCFs to be compared with exact quantum TCFs (without artificially damping in time), and to be obtained for non-linear operators which report on the dynamics of the entire delocalised quantum Boltzmann distribution. The almost perfect agreement between Matsubara and quantum  for the two highest bath strengths is strong evidence that the Matsubara dynamics hypothesis is correct: smoothing the imaginary-time Feynman paths to give a classical dynamics in an extended space correctly removes real-time coherences (which are evidently insignificant at the bath strengths considered).

The use of Matsubara dynamics also gives fresh physical insight into this old (system-bath) problem. It has long been known that the system-bath coupling enhances the effective strength of the `ring-polymer springs' along the system coordinate, causing delocalised static features, such as instantons, to become more compact.\cite{Litman2022,Litman2022a} The results of Sec.~IV show that the dynamics behaves analogously, with the system fluctuation modes (which correspond to the `spring' degrees of freedom) behaving much more harmonically than the centroid. Nonetheless, sufficient anharmonicity remains that a considerable number of fluctuation modes ($\sim 50$, in the calculations of Sec.~IV) need to be coupled to the dynamics of the centroid to give quantitative agreement with the exact quantum result.

This near-harmonic dynamics of the fluctuation modes is probably what made the Matsubara calculations numerically do-able. We found that we could analytically continue the dynamics into the complex plane (thus  converting the Matsubara phase into springs), where it remained numerically stable up to some maximum number of modes $M$ which increased with the bath coupling strength and decreased with the system anharmonicity. This suggests that Matsubara dynamics can probably be used as a practical  method for treating system-bath problems in the overdamped limit, where exact quantum methods such as HEOM\cite{Tanimura1989,Tanimura1991,Tanimura2020,Shi2011,Zhu2012} and QUAPI\cite{Makri1992,Topaler1994,Makri1995} (which are very efficient for treating relatively weak system-bath damping) are known to become expensive.  So far, we have treated dynamics on a single Born-Oppenheimer surface, but perhaps approaches such as that of ref.~\onlinecite{Chowdhury2021} can be used to extend the treatment developed here to non-adiabatic dynamics.

\begin{acknowledgments}
It is a pleasure to thank Jimin Li, Robert Jack, Ilija Srpak and William Moore for extensive discussions about the work in this paper.  AP acknowledges an EPSRC DTA PhD studentship from the UK Engineering and Physical Sciences Research Council; ESP acknowledges an External Research Studentship from Trinity College, Cambridge.
\end{acknowledgments}
\section*{AUTHOR DECLARATIONS
}
\subsection*{Conflict of Interest}
The authors have no conflicts to disclose.

\subsection*{Author Contributions}
\textbf{Adam Prada:} Conceptualisation (supporting), Formal analysis (equal), Investigation (lead), Methodology (lead), Software (lead), Visualisation (lead), Writing -- original draft (equal), Writing -- review \& editing (equal).
\textbf{Eszter S.~P\'os:} Conceptualisation (supporting), Investigation (supporting), Methodology (supporting), Writing -- review \& editing (supporting).
\textbf{Stuart C.~Althorpe:} Conceptualization (lead), Formal analysis (equal), Investigation (supporting), Supervision (lead), Writing -- original draft (equal), Writing -- review \& editing (equal).
\section*{DATA AVAILABILITY}
The data that support the findings of this study are available from the corresponding author upon reasonable request.

\appendix
\section{Potential of mean force}\label{app:mean_force}
Following ref.~\onlinecite{Trenins2018}, the potential of mean force is defined by
\begin{multline}
    \ue^{-\beta U_M(\vc{Q})} = \lim\limits_{N\to \infty} \left(\frac{m}{2\pi \beta_N \hbar^2}\right)^{(N-M)/2} N^{M/2}\\
    \int \ud \vc{q}\ \delta(\vc{q},\vc{Q}) \ue^{-\beta\left[W_N(\vc{q})-S_M(\vc{Q}))\right]}
    \label{eq:mf_potential}
\end{multline}
with
\begin{equation}
W_N(\vc{q}) = \frac{1}{N} \sum_{l=1}^{N} \left[\frac{m(q_l-q_{l-1})^2}{2\hbar^2 \beta^2/N^2 } + V(q_l)\right].
\label{eq:MF_W}
\end{equation}
and
\begin{equation}
S_M(\vc{Q}) = \sumM \frac{m \omega_n^2 Q^2}{2}.
\end{equation}
We evaluated $U_M(\vc{Q})$ on the fly by artificially increasing the frequencies of the $|n|>\mybar{M}$ modes using an adiabatic decoupling parameter $\gamma$ as described in ref.~\onlinecite{Trenins2018}.
\section{Equilibrium Matsubara GLE}\label{app:equilibrium}
In thermal equilibrium, the Kubo-transformed TCF is
\begin{multline}
    \widetilde{C}_{AB}^{\ [M]}(t) = \frac{1}{Z}\int\ud \vc{Q}\ \ud \vc{P}\ \ud \vc{X}\ \ud \vc{P}_{\mathrm{b}}\\
    \ue^{-\beta H_M\left(\vc{Q}, \vc{P}, \vc{X}, \vc{P}_{\mathrm{b}} \right)}\ue^{\ui\beta \theta\left(\vc{Q}, \vc{P}, \vc{X}, \vc{P}_{\mathrm{b}} \right)} A\left(\vc{Q}\right)B\left[\vc{Q}(t)\right].
    \label{eq:matsubara_tcf_system_bath}
\end{multline}
Following the procedure of Sec.~IIIA, we obtain 
\begin{equation}
    \begin{split}
        &H_M\left(\vc{Q}, \vc{P}, \vc{X}, \vc{{P}}_{\mathrm{b}}\right) -\ui \theta_\mathrm{B}\left(\vc{Q},\vc{P}\right) \to \\
        &= \sum_{n} \frac{P_n^2}{2m} + U_M(\vc{Q})
        + \sum_n \frac{|\omegan|\hat{\zeta}\bigl(|\omegan|\bigr)}{2}Q_n^2 \\
        & \quad + \sum_{\alpha,n}\left[\frac{{P}_{\alpha n}^{2}}{2\malpha} + \frac{\malpha \omega_{\alpha n}^2}{2}\left(X_{\alpha n} -\frac{\galpha Q_n}{\malpha \omega_{\alpha n}^2}\right)^2\right],
    \end{split}
    \label{eq:dist_analytic_cont_bath}
\end{equation}
where $\hat{\zeta}$ is the Laplace transform of the memory kernel
\begin{equation}
    \hat{\zeta}(s) = \sum_\alpha \frac{\galpha^2}{\malpha \omegaalpha^2}\frac{s}{s^2+\omegaalpha^2}.
\end{equation}
The analogous expressoin to \eqn{eq:noise_driving_dp} is then
\begin{multline}
    \sum_\alpha \galpha \biggl[\left(X_{\alpha n} -\frac{\galpha Q_n}{\malpha \omega_{\alpha n}^2} \right)\cos(\omegaalpha t)\\
    + \left(\frac{{P}_{\alpha n}}{\malpha \omegaalpha} + \ui  \frac{\omegan}{\omegaalpha} \left[X_{\alpha \mybar{n}} -\frac{\galpha Q_{\mybar{n}}}{\malpha \omega_{\alpha n}^2} \right]\right) \sin(\omegaalpha t)\biggr] \\ - \left[Q_n K_n(t) - \ui  Q_{\mybar{n}} L_n(t)\right].
    \label{eq:noise_driving}
\end{multline}
where the driving terms are
\begin{align}
    K_n(t) &= \omegan^2 \sum_\alpha \frac{\galpha^2}{\malpha \omegaalpha^2 \omega_{\alpha n}^2} \cos(\omegaalpha t)
\end{align}
and
\begin{align}
    L_n(t) &= \omegan \sum_\alpha \frac{\galpha^2}{\malpha \omegaalpha \omega_{\alpha n}^2} \sin(\omegaalpha t),
\end{align}
which give \eqnn{eq:K}{eq:L} of Sec.~\ref{sec:matsubara_gle}.
When $\partial K_n (t)/\partial t$ is defined, the two driving terms are related by\footnote{This is assuming that the derivative of $K_n(t)$ is defined, which seems always to be the case except  for a white noise bath at $t=0$.}
\begin{equation}
    L_n(t) = -\frac{1}{\omegan}\frac{\partial K_n (t)}{\partial t}.
\end{equation}
This gives rise to 
\begin{equation}
    \begin{split}
        m\ddot{Q}_n(t) = &-\frac{\partial U_M(\vc{Q}(t))}{\partial Q_n}  - \int_{0}^{t}\ud \tau\ \zeta(t-\tau)\dot{Q}_n(\tau)\\
        &\qquad + R_n^{\mathbb{C}}(t) - \left[Q_n K_n(t) - \ui  Q_{\mybar{n}} L_n(t)\right]
    \end{split},
    \label{eq:matsubara_gle}
\end{equation}
which is the equilibrium version of the Matsubara GLE (cf the non-equilibrium direct-product GLE of \eqn{eq:matsubara_gle_dp}). The noise term can be approximated to be real as described in Sec.~\ref{sec:decorrelation}. The driving terms \(K_n\) and \(L_n\) provide an additional memory, which means that even white noise simulations with a delta function memory kernel are no longer Markovian.

The  implementation of \eqn{eq:matsubara_gle} was found to have very similar numerical properties to its direct-product counterpart (\eqn{eq:matsubara_gle_dp}).
\section{Debye bath}
\label{app:baths}
The results of Sec.~IV used the Debye bath spectral density. Bath spectral densities are typically parametrised by the bath strength $\eta$ (coupling strength) and cut-off frequency $\omegac$, and are usually normalised, such that
\begin{equation}
\int_{0}^{\infty}\ud t\,\zeta(t) = \eta
\end{equation}
 Our calculations used the following well known or easily derivable expressions for the Debye bath:
\begin{align}
J(\omega) &= \eta \omega \frac{\omegac ^2}{\omegac ^2+\omega^2}\\
\aren  &= \eta \omegac \\
\zeta(t) &= \eta \omegac  \mathrm{e}^{-\omegac t}
\\
\hat{\zeta}(s) &=\frac{\eta \omegac}{\omegac + s}
\\
\omegaalpha &= \omegac \tan\left(\frac{\pi}{2}\frac{\alpha-1/2}{\nbath }\right),\ \alpha = 1, ..., \nbath
\\
w_\alpha &= \frac{\pi}{2} \frac{\eta \omegac }{\nbath }\frac{\omegaalpha}{J(\omegaalpha)}\\
\galpha &= \omegaalpha \sqrt{\frac{\eta \omegac  \malpha} {\nbath }}\\
\frac{\galpha^2}{\malpha \omegaalpha^2} &= \frac{\eta \omegac }{\nbath }\text{,\ \ for all }\alpha
\\
\galpha' &= \sqrt{\frac{\eta \omegac }{\beta \nbath}}
\\
K_n(t) &= \eta |\omegan| \omegac
\frac{\omegac \ue^{-|\omegan t|} - |\omegan|\ue^{-\omegac |t|}}{\omegac^2 - \omegan^2}
\\
L_n(t) &= \eta \omegan \omegac^2
\frac{\ue^{-|\omegan| t} - \ue^{-\omegac t}}{\omegac^2 - \omegan^2}
\end{align}
Corresponding expressions are easily derived for other spectral densities; see ref.~\onlinecite{Prada2022} for expressions for the white noise and Ohmic (with exponential cut-off) baths.  In general, $K_n$, $L_n$ and $\hat{\zeta}$ do not need to have closed analytic expressions, since they can easily be evaluated numerically. 

\section{Numerical integration of the analytically continued Matsubara GLE}\label{propagator}
To integrate the Matsubara GLE, we use the usual Liouvillian-splitting technique\cite{Tuckerman2010,Lawrence2019} to obtain the  propagator,
\begin{widetext}
\renewcommand*{\arraystretch}{2}
\begin{align}
    {P}_n &\gets {P}_n - \frac{\uDelta t^2}{4m}\zeta(0)\left({P}_n + \ui m \omegan Q_{\mybar{n}} \right)
    \\
        {P}_n &\gets {P}_n + \frac{\uDelta t}{2}\Biggl(-\diff{U_M(\vc{Q})}{Q_n} + m \omega_V^2 Q_n
        - \frac{\uDelta t}{m} \left[\sum_{j=1}^{i-1}\zeta(t_j) \Pi_n(t_{i-j}) + \frac{\zeta(t_i)\Pi_n(t_0)}{2}\right]
        + R^{\mathbb{C}}_n(t_i)
        - Q_n(0) \zeta(t_i)
         \Biggr)\label{eq:ac_int_force1}
    \\
    \begin{pmatrix} {P}_n \\ Q_n \\ P_{\mybar{n}} \\ Q_{\mybar{n}}\end{pmatrix} &\gets
    \begin{pmatrix}
        \cos(\omega_V \uDelta t) & - \frac{m \left(\omega_V^2 + \omegan^2\right)\sin (\omega_V \uDelta t)}{\omega_V} & - \ui  \frac{\omegan\sin (\omega_V \uDelta t)}{\omega_V} & 0
        \\
        \frac{\sin (\omega_V \uDelta t)}{m \omega_V} & \cos(\omega_V \uDelta t) & 0 &  \ui  \frac{\omegan\sin (\omega_V \uDelta t)}{\omega_V}
        \\
         \ui  \frac{\omegan\sin (\omega_V \uDelta t)}{\omega_V} & 0 & \cos(\omega_V \uDelta t) & - \frac{m \left(\omega_V^2 + \omegan^2\right)\sin (\omega_V \uDelta t)}{\omega_V}
        \\
        0 & - \ui \frac{ \omegan\sin (\omega_V \uDelta t)}{\omega_V} & \frac{\sin (\omega_V \uDelta t)}{m \omega_V} & \cos(\omega_V \uDelta t)
    \end{pmatrix}
    \begin{pmatrix} {P}_n \\ Q_n \\ P_{\mybar{n}} \\ Q_{\mybar{n}}\end{pmatrix}\label{eq:ref_sys_prop}
    \\
        {P}_n &\gets {P}_n + \frac{\uDelta t}{2}\Biggl(-\diff{U_M(\vc{Q})}{Q_n}+ m \omega_V^2 Q_n
        - \frac{\uDelta t}{m}\left[\sum_{j=1}^{i}\zeta(t_j) \Pi_n(t_{i+1-j}) + \frac{\zeta(t_{i+1})\Pi_n(t_0)}{2}\right]
         + R^{\mathbb{C}}_n(t_{i+1})
        - Q_n(0) \zeta(t_{i+1})\Biggr)
\label{eq:ac_int_force2}
    \\
    {P}_n &\gets {P}_n - \frac{\uDelta t^2}{4m}\zeta(0)\left({P}_n + \ui m \omegan Q_{\mybar{n}} \right)
\end{align}
\end{widetext}
in which $m\omega_V^2q^2/2$ is the harmonic part of $V(q)$ and $\Pi_n(t) = {P}_n(t) + \ui m \omegan Q_{\mybar{n}}(t)$. \Eqn{eq:ref_sys_prop} is a generalisation of the ``ring-polymer reference system propagator'' of ref.~\onlinecite{Ceriotti2010} to integrate the analytically continued equations of motion
\begin{align}
    \dot{Q}_n &= \frac{{P}_n}{m} + \ui \omegan Q_{\mybar{n}}\\
    \dot{{P}}_n &= -m \left(\omega_V^2 + \omegan^2\right) Q_n - \ui \omegan {P}_{\mybar{n}}\\
    \dot{Q}_{\mybar{n}} &= \frac{{P}_{\mybar{n}}}{m} + \ui \omega_{\mybar{n}} Q_n\\
    \dot{{P}}_{\mybar{n}} &= -m \left(\omega_V^2 + \omega_{\mybar{n}}^2\right) Q_{\mybar{n}} - \ui \omega_{\mybar{n}} {P}_n
\end{align}
To propagate solutions to the equilibrium version of the Matsubara GLE of \eqn{eq:matsubara_gle}, one  replaces the driving term $Q_n(0)\zeta(t)$ in \eqnn{eq:ac_int_force1}{eq:ac_int_force2} with $\left[Q_n(0) K_n(t) - \ui  Q_{\mybar{n}(0)} L_n(t)\right]$.

\clearpage

\section{Harmonic correction}\label{app:harmonic_correction}
To treat the modes $\mybar{M}<|n|\le \mybar{M_\text{eff}}$ harmonically, we decompose the Matsubara TCF as
\begin{align}
    \widetilde{C}_{q^2 q^2}(t) &= \sum_{n_1,n_2}^{\{M_{\mathrm{eff}}\}} \langle Q_{n_1}^2 Q_{n_2}^2(t) \rangle
    \\
    &= \sum_n^{\{M_{\mathrm{eff}}\}} \langle Q_{n}^2 Q_{n}^2(t) \rangle + \sum_{n \neq 0}^{\{M_{\mathrm{eff}}\}} \langle Q_{n}^2 Q_{\mybar{n}}^2(t) \rangle\nonumber \\&\quad+ \sum_{n_1 \neq \pm n_2}^{\{M_{\mathrm{eff}}\}} \langle Q_{n_1}^2 Q_{n_2}^2(t) \rangle
\end{align}
where  $\sum^{\{M_{\mathrm{eff}}\}}$ denotes a sum over all the system Matsubara modes $|n|\le \mybar{M_\text{eff}}$. We then further split the sums into
\begin{align}
    \begin{split}
        &\widetilde{C}_{q^2 q^2}(t) =\\
        &= \underbrace{\sum_n^{\{M\}} \langle Q_{n}^2 Q_{n}^2(t) \rangle + \sum_{n \neq 0}^{\{M\}} \langle Q_{n}^2 Q_{\mybar{n}}^2(t) \rangle + \sum_{n_1}^{\{M\}} \sum_{n_2 \neq \pm n_2}^{\{M\}} \langle Q_{n_1}^2 Q_{n_2}^2(t) \rangle}_{\displaystyle=\langle \hat{q}^2 \hat{q}^2(t)\rangle_{M}}
        \\
        &+\underbrace{\sum_n^{\{M'\}} \langle Q_{n}^2 Q_{n}^2(t) \rangle + \sum_{n \neq 0}^{\{M'\}} \langle Q_{n}^2 Q_{\mybar{n}}^2(t) \rangle + \sum_{n_1}^{\{M'\}} \sum_{n_2 \neq \pm n_2}^{\{M'\}} \langle Q_{n_1}^2 Q_{n_2}^2(t) \rangle}_{\displaystyle  \textrm{ purely harmonic}}
        \\
        &+ \underbrace{\sum_{n_1}^{\{M\}} \sum_{n_2}^{\{M'\}} \langle Q_{n_1}^2 Q_{n_2}^2(t) \rangle + \sum_{n_1}^{\{M'\}} \sum_{n_2}^{\{M\}} \langle Q_{n_1}^2 Q_{n_2}^2(t) \rangle.}_{\displaystyle \textrm{mixed}}
    \end{split}
\end{align}
where $\sum^{\{M\}}$ denotes a sum over the  $|n|\le \mybar{M}$ modes which are included in the full anharmonic simulation, and $\sum^{\{M'\}}$ a sum over the modes $\mybar{M}<|n|\le \mybar{M_\text{eff}}$ which are treated harmonically. The purely harmonic terms are evaluated accurately from numerical propagations carried out independently for each $n$, using $V(q)\simeq m\omega_V^2q^2/2$. The mixed terms are calculated using 
\begin{align}
    \langle Q_{n_1}^2 Q_{n_2}^2(t) \rangle \simeq \langle Q_{n_1}^2 \rangle \langle Q_{n_2}^2(t) \rangle \textrm{ for } n_1 \neq \pm n_2
    \label{eq:decorrelation}
\end{align}
which is exact if the modes $n_2$ are harmonic.
For sufficiently high $|n|$ (typically $|n|\gtrsim 200$) the difference between the direct-product and equilibrium values was found to be negligible, allowing the $\langle Q_{n_2}^2(t) \rangle$ terms to be replaced by their equilibrium values
\begin{equation}
    \langle Q_n^2 \rangle = \frac{1}{\beta m \left(\omega^2 + \omegan^2 + \frac{|\omegan|}{m}\hat{\zeta}(|\omegan|)\right)}.
\end{equation}
\begin{figure}
    \includegraphics{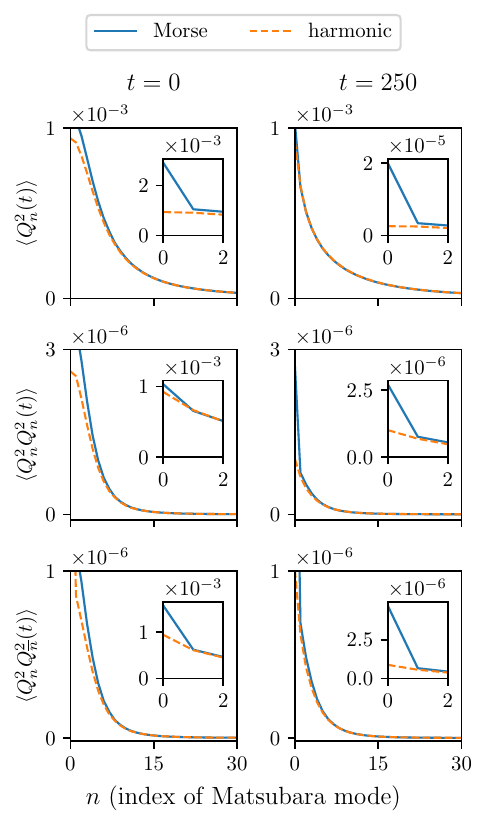}
    \caption{Expectation values $\langle Q^2_n(t)\rangle$ and time-correlation functions $\langle Q^2_n Q^2_n(t)\rangle$ and $\langle Q^2_n Q^2_{\mybar{n}}(t)\rangle$, obtained by decomposing the real-noise Matsubara calculations of $\langle q^2 q^2(t)\rangle$ (using the parameters of Table~\ref{tab:1}) into contributions from individual modes $Q_n$, compared with the purely harmonic results. All units are in a.u.}
    \label{fig:9cases}
\end{figure}

\bibliography{article}

\end{document}